%% file: ines.tex
\newcommand{\ONECOLUMN}{}

\ifdefined\ONECOLUMN
	\documentclass[review]{elsarticle}
\else
	\documentclass[10pt,twocolumn,twoside]{IEEEtran}
\fi

\usepackage{amsmath,amsfonts,amssymb,latexsym,epsfig,mathtools,appendix}
\usepackage[active]{srcltx}

\usepackage{ulem}

\usepackage{algorithm}
\usepackage[noend]{algorithmic}
\makeatletter
\newcounter{ALC@tempcntr}% Temporary counter for storage

\makeatother

\usepackage{xcolor}

\usepackage{subcaption}

\usepackage{graphicx}

\usepackage{tikz}
\usetikzlibrary{calc,positioning,intersections,decorations}
\usepackage{tkz-euclide}
\usetkzobj{all}

\usepackage{tabu}

\newcommand{\dfn}{\triangleq} 

\newcommand{\rw}{\rightarrow}

\newcommand{\mX}{{\mathcal X}}

\newcommand{\mK}{{\mathcal K}}
\newcommand{\mT}{{\mathcal T}}
\newcommand{\Real}{\mathbb{R}}
\newcommand{\sS}{{\sf S}}

\newcommand{\sd}{{\sf d}}
\newcommand{\sw}{{\sf w}}

\newcommand{\mbE}{\mathbb{E}}

\newcommand{\bx}{{\bf x}}
\newcommand{\by}{{\bf y}}

\newtheorem{teorema}{\bf Theorem}
\newtheorem{lema}{\bf Lemma}

\newtheorem{hipotesis}{\bf Assumption}
\newtheorem{nota}{\bf Remark}

\input{manu_preamble}

\begin{document}

\begin{frontmatter}

\title{Analysis of a nonlinear importance sampling scheme for Bayesian parameter estimation in state-space models}

\author{Joaqu\'in M\'iguez$^\dagger$}
\ead{joaquin.miguez@uc3m.es}

\author{In\'es P. Mari\~no$^\star$}
\ead{ines.perez@urjc.es}

\author{Manuel A. V\'azquez$^\dagger$}
\ead{mvazquez@tsc.uc3m.es}

\address{$^\dagger$Department of Signal Theory \& Communications, Universidad Carlos III de Madrid. Avenida de la Universidad 30, 28911 Legan\'es, Madrid, Spain.}

\address{$^\star$Department of Biology and Geology, Physics and Inorganic Chemistry, Universidad Rey Juan Carlos. C/ Tulip\'an s/n, 28933 M\'ostoles, Madrid, Spain.}

%\maketitle

\begin{abstract}
The Bayesian estimation of the unknown parameters of state-space (dynamical) systems has received considerable attention over the past decade, with a handful of powerful algorithms being introduced. In this paper we tackle the theoretical analysis of the recently proposed {\it nonlinear} population Monte Carlo (NPMC). This is an iterative importance sampling scheme whose key features, compared to conventional importance samplers, are (i) the approximate computation of the importance weights (IWs) assigned to the Monte Carlo samples and (ii) the nonlinear transformation of these IWs in order to prevent the degeneracy problem that flaws the performance of conventional importance samplers. The contribution of the present paper is a rigorous proof of convergence of the nonlinear IS (NIS) scheme as the number of Monte Carlo samples, $M$, increases. Our analysis reveals that the NIS approximation errors converge to 0 almost surely and with the optimal Monte Carlo rate of $M^{-\frac{1}{2}}$. Moreover, we prove that this is achieved even when the mean estimation error of the IWs remains constant, a property that has been termed {\it exact approximation} in the Markov chain Monte Carlo literature. We illustrate these theoretical results by means of a computer simulation example involving the estimation of the parameters of a state-space model typically used for target tracking.
\end{abstract}

\begin{keyword}
Importance sampling; population Monte Carlo; state space models; Bayesian inference; adaptive importance sampling; parameter estimation.
\end{keyword}

\end{frontmatter}

% INTRODUCTION
% --------------------------------------
\section{Introduction}
\label{sec:intro}

The estimation of the static unknown parameters of state-space dynamic models is a classical problem in statistical signal processing \cite{Jansson96,Storvik02,Andrieu03,Ding10,Ding13,Kokkala15} which has also received considerable attention, very recently, from the computational statistics community \cite{Andrieu10,Koblents15,Crisan15} (see also \cite{Kantas15} for a recent survey) partly because of the ubiquity of the problem in science and engineering and partly because of the availability of more powerful computational resources to address it.

The particle Markov chain Monte Carlo (pMCMC) method originally proposed in \cite{Andrieu10} has been rapidly adopted by researchers in signal processing \cite{Olsson11,Vu14,Kokkala15,Kwon16,Ala16}. This is a Markov chain Monte Carlo (MCMC) algorithm \cite{Fitzgerald01} where the target probability density function (pdf) is the posterior density of the unknown parameters conditional on the available observations. This pdf is analytically intractable and, hence, it is approximated (for each element of the chain) via particle filtering \cite{Gordon93,Doucet01b,Doucet00,Djuric03,Cappe07}. The most popular MCMC schemes (including Metropolis and Metropolis-Hastings algorithms) admit a pMCMC implementation. A key feature of these methods is that they have the so-called {\it exact approximation} property. This means that, even if the acceptance test of the MCMC algorithm is only approximate (since the true target pdf is intractable), the stationary distribution of the Markov chain is still actual posterior density of the parameters. While popular, pMCMC procedures suffer from the same limitations as regular MCMC schemes \cite{Fitzgerald01,Robert04}:
\begin{itemize}
\item Convergence of the chain is purely asymptotic (no convergence rates are known) and potentially very slow (a problem made worse by the particle approximation).
\item The Monte Carlo samples in the chain are correlated, which reduces the accuracy of estimators compared to methods that produce independent samples.
\item If the target pdf is multimodal, MCMC algorithms may get trapped in local maxima of the function.
\end{itemize}

An alternative to pMCMC methods is to employ schemes based on importance sampling (IS) \cite{Robert04}. This class of techniques includes population Monte Carlo (PMC) \cite{Cappe04}, the sequential Monte Carlo square (SMC$^2$) of \cite{Chopin12} or the nested particle filter of \cite{Crisan15}. PMC is an iterative IS scheme in which the proposed functions used to generate Monte Carlo samples (and, hence, to approximate the posterior probability distribution of the unknown parameters) are improved across the iterations of the algorithm. See \cite{Hong10,Martino15,Bugallo15,Elvira17} for recent applications, and new developments, of this methodology in statistical signal processing. SMC$^2$ is a generalisation of the iterative batch importance sampling (IBIS) algorithm of \cite{Chopin02}. It mimics the standard particle filter, but the Monte Carlo samples are drawn from the space of the (static) parameters and they are sequentially updated using a pMCMC kernel. All these methods, including SMC$^2$, are batch, meaning that the whole record of observations is typically processed many times. A purely recursive version of the SMC$^2$ algorithm has been proposed in \cite{Crisan15}. The reduction in computational complexity, however, is obtained at the expense of a reduction in the convergence rate of the algorithm. It is worth mentioning that all these techniques (including pMCMC) can be fit within the theoretical framework of sequential Monte Carlo samplers introduced in \cite{DelMoral06}.

The key feature of IS-based methods is that the Monte Carlo samples (used to approximate the target distribution) are generated from almost-arbitrary proposal functions and then assigned importance weights (IWs). While this is a very flexible approach, it suffers from the well-known problem of degeneracy of IWs \cite{Kong94,Doucet00,Robert04,Koblents15}: when the target pdf is concentrated in a very small region of the space of the unknowns, the largest IW tends to be orders of magnitude greater than all other IWs. As a result the IS-based scheme practically yields a degenerate one-sample approximation.

In this paper we address the analysis of the nonlinear population Monte Carlo (NPMC) algorithm proposed in \cite{Koblents15}. In the latter scheme, the IWs undergo a nonlinear transformation to control their variance and, in this way, mitigate the degeneracy problem. In \cite{Koblents15} it was proved that the approximation of the target distribution produced at each iteration of the NPMC method converges asymptotically, with the number of Monte Carlo samples $M$, and almost surely (a.s.). Therefore, the weight transformation preserves asymptotic convergence, while it has been shown through numerical examples that performance for finite $M$ is consistently improved compared to conventional PMC procedures. The analysis in \cite{Koblents15}, however
\begin{itemize}
\item relies on the exact computation of the IWs, which is not feasible for general state-space models,
\item and does not provide explicit convergence rates\footnote{Error rates are found in \cite{Koblents15} for convergence in probability (not for almost sure convergence) when the IWs are computed exactly.}
\end{itemize}

In this paper we analyse the performance of NPMC methods for the Bayesian estimation the unknown parameters of state space models. Based on some unbiasedness properties of particle filters, we prove that IS with nonlinearly-transformed IWs also yields asymptotic convergence when the weights are approximate, i.e., computed via a particle filter with a fixed computational budget that introduces non-vanishing errors. In other words, we prove that the nonlinear importance sampler enjoys the same exact approximation property as pMCMC and SMC$^2$ algorithms. Moreover, the analysis of this paper also extends considerably the results of \cite{Koblents15} by obtaining an explicit (and almost sure) estimation error rate of order $M^{-\frac{1}{2}+\epsilon}$, where $\epsilon>0$ is an arbitrarily small constant. This result holds for approximate weights and under mild assumptions typical of classic IS analyses. It is worth mentioning that the analytical approach developed in this paper can be applied, in a rather natural way, to the study of recently proposed PMC-like algorithms \cite{Martino15,Elvira15b} when the target distribution is the posterior density of the parameters of a state space model.

The rest of the paper is organised as follows. The necessary background material, including notation, state-space models and particle filters, is presented in Section \ref{ssStateSpace}. The nonlinear IS scheme and its iterative implementation (the NPMC algorithm) are detailed in Section \ref{sNPMC} for the case in which the target probability distribution is the posterior distribution of the unknown parameters of a state-space model. In Section \ref{sConvergence} we introduce the new analytical results on the convergence of nonlinear importance samplers, which is the main contribution of the paper. We illustrate the exact approximation property, and numerically compare the NPMC algorithm with a pMCMC scheme through computer simulations for a target tracking model in Section \ref{sec:simulations}. Finally, some brief concluding remarks are made in Section \ref{sConclusions}.

\section{Background and problem statement} \label{ssStateSpace}
\subsection{State-space model}
A Markov state-space model consists of two sequences of random variables (r.v.'s), $\{\bx_n\}_{n\ge 0}$ and $\{\by_n\}_{n \ge 1}$. The first sequence, $\{\bx_n\}$, is termed the system state. We assume it takes values on some space $\mX \subseteq \Real^{d_x}$, hence $\bx_n$ is a random $d_x \times 1$ vector. The state dynamics are described by a prior probability measure $\mK_0(\sd\bx_0)$ and a sequence of Markov kernels $\mK_{n,\theta}(\sd\bx_n|\bx_{n-1})$ that depend on a parameter vector $\theta \in {\sf S} \subset \Real^{d_\theta}$. In this paper, $\theta$ is assumed unknown and modelled as a random vector, with prior pdf $p_0(\theta)$ with respect to (w.r.t.) the Lebesgue measure. The support set of the parameter vector, ${\sf S}$, is assumed to be compact.

The state $\bx_n$ cannot be observed directly. Instead, some noisy observations $\by_n \in \mathcal Y \subseteq \Real^{d_y}$, $n=1,2,\ldots$, are collected. We note that $\by_n$ is a $d_y \times 1$ vector, with $d_y \neq d_x$ in general.

We assume that the observations are conditionally independent given the system states and the parameter vector $\theta$, with a conditional pdf w.r.t.  the Lebesgue measure, denoted $l_{n,\theta}(\by_n|\bx_n) > 0$, which depends on the parameter vector $\theta$ as well. 

%%%%%%%%%%%%%%
\subsection{The optimal filter and its Monte Carlo approximation} \label{ssFilter}

Let $\by_{1:n}=\{\by_1,\ldots,\by_n\}$ denote the sequence of observations collected up the time $n$. The posterior probability measure of the state $\bx_n$ conditional on the observations  $\by_{1:n}$ and the parameter vector $\theta$ is denoted $\pi_{n,\theta}$, i.e., for any Borel set $A \subset \mX$,
\begin{equation}
\pi_{n,\theta}(A)=\int_A \pi_{n,\theta}(\sd \bx)
\end{equation}
is the posterior probability of the event ``$\bx_n \in A$'', given $\theta$ and  $\by_{1:n}$. 

Similarly, $\xi_{n,\theta}$ denotes the posterior probability measure of $\bx_n$ conditional on $\theta$ and $\by_{1:n-1}$ (i.e., not including $\by_n$). This is often referred to as the one-step-ahead predictive measure (\cite{Bain08}, Chapter 10). For a Borel set  
$A \subset \mX$,
\begin{equation}
\xi_{n,\theta}(A)=\int_A \xi_{n}(\sd \bx)
\end{equation}
is the posterior probability of the event ``$\bx_n \in A$'', given $\theta$ and $\by_{1:n-1}$.

We refer to $\pi_{n,\theta}$ as the optimal filter conditional on the parameter vector $\theta$. It is not possible, in general, to obtain either  $\pi_{n,\theta}$  or $\xi_{n,\theta}$ in closed-form (with the notable exception of linear-Gaussian state space models, for which 
$\pi_{n,\theta}$  and  $\xi_{n,\theta}$ are computed recursively and exactly using the Kalman flter \cite{Anderson79}) and, therefore, 
numerical approximation algorithms are needed. One of the most popular schemes is the standard particle filter, also known as bootstrap filter (BF) \cite{Gordon93,Kitagawa96,Doucet00}.

The BF with $N$ particles (i.e., Monte Carlo samples on the state space $\mX$) conditional on a given parameter vector $\theta$  can be briefly outlined as follows.
\begin{enumerate}
\item \textit{\textbf{Initialisation.}} Draw $N$ samples $\bx_0^1, \ldots, \bx_0^N$ from the prior distribution $\mK(\sd\bx_0)$. The particle approximation of $\pi_{0,\theta}(\sd\bx_0) \equiv \mK_0(\sd\bx_0)$ is
\begin{equation}
\pi_{0,\theta}^N(\sd\bx_0) = \frac{1}{N} \sum_{i=1}^N \delta_{\bx_0^i}(\sd\bx_0),
\end{equation}
where $\delta_{\bx_0^i}$ denotes the Dirac delta measure centred at $\bx_0^i \in \mX$.

\item \textbf{\textit{Recursive step.}} Given the approximation $\pi_{n-1,\theta}^N(\sd\bx_{n-1}) =  \frac{1}{N} \sum_{i=1}^N \delta_{\bx_{n-1}^i}(\sd\bx_{n-1})$, take the following steps:
	\begin{enumerate}
	\item Randomly propagate each particle using the Markov kernel in the model, i.e., draw $\tilde \bx_n^i$ from $\mK_{n,\theta}(\sd\bx_n|\bx_{n-1}^i)$, $i = 1, ..., N$.
	\item Compute IWs, $\tilde u_n^i = l_{n,\theta}(\by_n|\tilde \bx_n^i)$, for $i=1, ..., N$, and
	\item normalise them as
	\begin{equation}
	u_n^i = \frac{ \tilde u_n^i }{ \sum_{j=1}^N \tilde u_n^j }, \quad i=1, ..., N.
	\end{equation}
	\item Resample: draw $N$ times independently from the discrete distribution $\tilde \pi_{n,\theta}^N(\sd\bx_n)  =  \sum_{i=1}^N u_n^i\delta_{\tilde \bx_n^i}(\sd\bx_n)$ and denote the resulting samples as $\{ \bx_n^i \}_{i=1}^N$. Construct the  {\it unweighted} approximation $\pi_{n,\theta}^N(\sd\bx_n) = \frac{1}{N} \sum_{i=1}^N \delta_{\bx_n^i}(\sd\bx_n)$.
	\end{enumerate}
\end{enumerate}

The resampling step (d) above can be implemented in a number of different ways (see, e.g., \cite{Douc05,Bain08} or \cite{Cappe07} for a brief survey of methods). Here, for simplicity, we have adopted a scheme which is often referred to as multinomial resampling \cite{Doucet00,Douc05} but most asymptotic convergence results hold true for several other schemes as well \cite{DelMoral04,Bain08}. The measure-valued r.v.  $\pi_{n,\theta}^N$ is an approximation of the optimal filter $\pi_{n,\theta}$ (conditional on $\theta$). Let us use the shorthand
\begin{equation}
(f,\pi)=\int f(\bx)\pi(\sd \bx)
\end{equation}
for the integral of a real function $f:{\Real}^d \rightarrow {\Real}$ w.r.t. a measure $\pi$. Under very mild assumptions it can be shown that 
\begin{equation}
\lim_{N\rightarrow \infty}(f,\pi_{n,\theta}^N)=(f,\pi_{n,\theta})
\end{equation}
almost surely (a.s.) for any bounded function $f:\mX \rightarrow {\Real}$ \cite{DelMoral04,Bain08}. Moreover, if we denote $||f||_\infty=\sup |f(\bx)|$, $E[Z]$ indicates the expected value of a r.v. $Z$ and $||Z||_p=(E[|Z|^p])^\frac{1}{p}$ is its $L_p$ norm ($p\ge 1$), then it can be proved \cite{Miguez13b} that
\begin{equation}
||(f,\pi_{n,\theta}^N)-(f,\pi_{n,\theta})||_p \le \frac{C_n ||f||_\infty}{\sqrt{N}}
\end{equation}
where $C_n$ is a constant independent of $N$ and
\begin{equation}
(f,\pi_{n,\theta}^N)=\int f(\bx_n) \pi_{n,\theta}^N(\sd \bx_n)=\frac{1}{N}\sum_{i=1}^N f(\bx_n^i).
\end{equation}

The algorithm also produces a Monte Carlo approximation of the predictive measure $\xi_{n,\theta}$, namely
\begin{equation}
\xi_{n,\theta}^N(\sd\bx_n) = \frac{1}{N}\sum_{i=1}^N \delta_{\tilde \bx_n^i}(\sd\bx_n).
\end{equation} 
If we write $\by=\by_{1:n}$ for the complete sequence of observations up to time $n$, it turns out that the conditional pdf of $\by$ given  the parameter vector $\theta$, denoted $\ell(\by|\theta)$, can  be written in terms of integrals w.r.t. to the predictive  measures   $\xi_{k,\theta}$, $k=1,\ldots,n$. To be specific,
\begin{equation}
\ell(\by|\theta) = \prod_{k=1}^n (l_{k,\theta}(\by_k|\cdot),\xi_{k,\theta}),
\label{eqDefEll}
\end{equation}
where
\begin{equation}
(l_{k,\theta}(\by_k|\cdot),\xi_{k,\theta})=\int_\mX l_{k,\theta}(\by_k|\bx_k) \xi_{k,\theta}(\sd \bx_k).
\end{equation}

The conditional pdf $\ell(\by|\theta)$ is the likelihood of the parameter vector $\theta$ given the available data $\by$ and the BF yields 
 the straightforward estimator
\begin{equation}
\ell^N(\by|\theta) = \prod_{k=1}^n (l_{k,\theta}(\by_k|\cdot),\xi_{k,\theta}^N) 
\label{eqUBEstimator} 
\end{equation}
which can be shown to be unbiased (i.e., $E[\ell^N(\by|\theta)]=\ell(\by|\theta)$) under very mild assumptions (\cite{DelMoral04}, Theorem 7.4.2).

\subsection{Problem statement}

Let $\by_=\{\by_1,\ldots,\by_R\}$ be the available data set, with $R<\infty$. Our goal is to approximate the probability measure associated to the posterior pdf of the parameter vector, $\theta$, given the data, $\by$. We denote this pdf as $p(\theta|\by)$ and it is straightforward to show, using Bayes' theorem, that
\begin{equation}
p(\theta|\by) \propto \ell(\by|\theta) p_0(\theta)
\end{equation}
where, we recall, $p_0(\theta)$ is the prior pdf of $\theta$.

In the next section, we describe an iterative importance sampling algorithm, originally introduced in \cite{Koblents15}, for the approximation of $p(\theta|\by)\sd \theta$.
%%%%%%
%
%%%%%%
\section{Algorithm} \label{sNPMC}

The NPMC algorithm of \cite{Koblents15} is an iterative importance sampling (IS) scheme that seeks to approximate a target probability distribution, in our case given by the posterior pdf $p(\theta|\by)$, using weighted Monte Carlo samples. It generates a sequence of proposal pdf's $q_k(\theta)$, $k = 1, \ldots, K$, from which samples can be drawn and importance weights (IWs) can be computed. This sequence of proposals is expected to yield increasingly better approximations of the target as the algorithm converges. The key feature of the NPMC method, which departs from the classical PMC technique of \cite{Cappe04}, is to compute a set of {\it transformed} importance weights (TIWs) by applying a nonlinear function to the standard IWs. The aim of this transformation is to mitigate the well-known problem of the degeneracy of the IWs (common to many IS methods, see \cite{Doucet00,Koblents15}) by controlling the weight variability.

For the case of general state space models, an additional difficulty encountered when trying to estimate the unknown model parameters (denoted $\theta$ in our setup) is that the likelihood $\ell(\by|\theta)$ is intractable. In the last few years, though, it has become a common approach to approximate this likelihood via particle filtering (PF) (see, e.g., \cite{Koblents15,Andrieu10,Andrieu09,Chopin12}). To be specific, we let $\ell^N(\by|\theta)$ stand for the approximation of $\ell(\by|\theta)$ computed using a standard bootstrap filter (BF) \cite{Gordon93,Doucet01} with $N$ particles (see equation (\ref{eqUBEstimator}) in Section  \ref{ssFilter}). One key feature of this approach, that we exploit for our analysis in Section \ref{sAnalysis}, is that $\ell^N(\by|\theta)$ can be proved to be an unbiased estimator of $\ell(\by|\theta)$ \cite{DelMoral04,Crisan15par}. 
 
The NPMC algorithm applied to a state space model, with $K$ iterations, $M$ Monte Carlo samples per iteration, plain Gaussian proposals $\{ q_k \}_{k\ge1}$, and approximate likelihoods is outlined below.

\noindent \textit{\textbf{Initialisation}}. Draw $M$ i.i.d. samples $\theta_0^1, \theta_0^2, \ldots, \theta_0^M$ from the prior pdf $p_0(\theta)$. Then,
\begin{enumerate}
	\item compute non-normalised IWs $\tilde \sw_0^i \propto \ell^N(\by|\theta_0^i)$, $i=1, ..., M$, 
	\item compute TIWs as $\hat \sw_0^i = \mT_M\left(i, \{\tilde \sw_0^j\}_{j=1}^M\right)$, where $\mT_M : \{1, \ldots, M\} \times \{ \tilde \sw_0^j \}_{j=1}^M \rw [0, +\infty)$ is a nonlinear transformation, and  
	\item normalise the TIWs, $\sw_0^i = \frac{\hat \sw_0^i}{\sum_{j=1}^M \hat \sw_0^j}$, $i=1, ..., M$.
\end{enumerate}

\noindent \textit{\textbf{Iteration}}. For $k = 1, \ldots, K$, take the following steps:
\begin{enumerate}
\item Let $q_k(\theta)={\cal N}(\theta|\mu_k,\Sigma_k)$ be a multivariate Gaussian pdf with mean vector and covariance matrix obtained, respectively, as
\begin{equation}
\mu_k = \sum_{i=1}^M \sw_{k-1}^i \theta_{k-1}^i
\quad \mbox{and} \quad 
\Sigma_k = \sum_{i=1}^M \sw_{k-1}^i \left(
	\theta_{k-1}^i - \mu_k
\right)\left(
	\theta_{k-1}^i - \mu_k
\right)^\top.
\end{equation}
Note that the random variates $\theta_{k-1}^i$, $i=1, ..., M$, are $d_\theta \times 1$ vectors. The superscript ${}^\top$ denotes transposition.

\item Draw i.i.d. samples $\theta_k^i$, $i=1, ..., M$,  from $q_k(\theta)$.
\item Compute IWs, $\tilde \sw_k^i = \frac{ \ell^N(\by|\theta_k^i) p_0(\theta_k^i) }{ q_k(\theta_k^i) }$, $i=1, ..., M$.
\item Compute TIWs, $\hat \sw_k^i = \mT_M\left( i,\{\tilde \sw_k^j\}_{j=1}^M \right)$, $i=1, ..., M$, using the same nonlinear map as for $k=0$.
\item Normalise the TIWs, $\sw_k^i = \frac{\hat \sw_k^i}{\sum_{j=1}^M \hat \sw_k^j}$, $i=1, ..., M$.
\end{enumerate}

Following \cite{Koblents15}, the nonlinear map $\mT_M$ of choice is a ``clipping'' transformation. In particular, let $i_1, i_2, ..., i_M$ be a permutation of the indices $1, 2, ..., M$ such that the IWs become ordered, namely $\tilde \sw_k^{i_1} \ge \tilde \sw_k^{i_2} \ge \cdots \ge \tilde \sw_k^{i_M}$. The clipping transformation $\mT_M$, with parameter $1 \le M_c \le \sqrt{M}$, flattens the $M_c$ largest IWs and makes them equal to the $M_c$-th non-normalised IW, $\tilde \sw_k^{i_{M_c}}$. Specifically, for each $j=1, ..., M$, we obtain
\begin{equation}
\hat \sw_k^j = \mT_M\left( j,\{\tilde \sw_k^l\}_{l=1}^M \right) = \left\{
	\begin{array}{ll}
		\tilde \sw_k^{i_{M_c}}, &\mbox{if $\tilde \sw_k^j \ge \tilde \sw_k^{i_{M_c}}$},\\
		\tilde \sw_k^j, &\mbox{if $\tilde \sw_k^j < \tilde \sw_k^{i_{M_c}}$},\\
	\end{array}
\right..
\label{eqClipping}
\end{equation}
Other choices of $\mT_M$ are possible (e.g., tempering schemes) but clipping has been found particularly effective in practice \cite{Koblents15}. The choice of Gaussian proposals (in step 1 of the {\it Iteration}) is made merely for simplicity. Other (more efficient) possibilities exist, but we stick to this formulation as it is sufficient for the purpose of this paper. 

Given $A  \subseteq \sS$, being $\sS$ the support set of the parameter vector $\theta$ described in Section \ref{ssStateSpace},
let $\mu_\by(A)=\int_A p(\theta|\by)\sd\theta$ denote the posterior probability measure (conditional on the observed data $\by$) associated to the parameter vector $\theta$. This measure yields the full probabilistic description of $\theta$ given the available observations. If  $\mu_\by$ is available, then we can compute various types of estimators and assess the associated errors. For example, the posterior-mean estimator is 
\begin{equation}
\hat \theta_* = \int_\sS \theta \mu_\by(\sd\theta),
\end{equation}
and it minimises the mean square error (MSE). For an arbitrary estimator $\hat \theta$, the MSE can also be written as an integral w.r.t. $\mu_\by(\sd\theta)$, namely, 
\begin{equation}
\mbox{MSE}(\hat \theta) = \int_\sS  (\theta - \hat \theta)^2 \mu_\by(\sd\theta).
\end{equation}

The proposed NPMC algorithm yields a sequence of importance sampling (i.e., weighted Monte Carlo) approximations of $\mu_\by(\sd\theta)$. To be specific, at each iteration $k$ we obtain the random probability measure
\begin{equation}
\mu_{\by,k}^M(\sd\theta) = \sum_{i=1}^M \sw_k^i \delta_{\theta_k^i}(\sd\theta),
\end{equation} 
where $\delta_{\theta_k^i}$ denotes the Dirac delta measure centred at $\theta_k^i$. Using $\mu_{\by,k}^M(\sd\theta)$ we can approximate any parameter estimator. For instance,  
$
 \hat \theta^M_k = \sum_{i=1}^M \sw_k^i \theta_k^i
$ 
is the approximation of the posterior mean $\hat \theta_*$. The corresponding minimum MSE can also be approximately computed as
\begin{equation}
\mbox{MSE}(\hat \theta^M_k) = \sum_{i=1}^M \sw_k^i \| \theta_k^i - \hat \theta^M_k \|^2.
\end{equation}

In the next section we analyse the convergence of the approximate measure $\mu_{\by,k}^M$ as $M\rightarrow\infty$ in a single iteration (i.e., for a given $k$) when the number of particles $N$ used to approximate the likelihood via the BF (i.e., the estimate  $\ell^N(\by|\theta)$ of  $\ell(\by|\theta)$) is kept constant and finite.

%%%%%%%
%
%%%%%%%
\section{Analysis} \label{sAnalysis} \label{sConvergence}

Consider a single iteration $k$ in the NPMC algorithm, with a fixed importance density $q_k \equiv q$. We refer to the random measure $\mu_{\by,k}^M(\sd\theta) = \sum_{i=1}^M \sw_k^i \delta_{\theta_k^i}(\sd\theta)$ computed via the TIWs $\sw_k^i$, $i=1, .., M$, as a nonlinear importance sampling (NIS) approximation of $\mu_\by(\sd\theta)$. Our aim in this section is to assess whether $\mu_{\by,k}^M(\sd\theta)$ converges towards the true measure $\mu_\by(\sd\theta)$ or not as $M\rw\infty$. To do this, there are two issues that need to be handled and make the analysis more difficult compared to a conventional IS method (that relies on the standard IWs, rather than the TIWs). These issues are:
\begin{itemize}
\item[(i)] the distortion in the Monte Carlo approximation due to the clipping of the weights, which  introduces additional  bias (compared to the use of standard IWs); and
\item[(ii)] the impossibility to compute the IWs, and hence the TIWs, exactly, since the likelihood $\ell(\by|\theta)$ is intractable and we work with the particle approximation  $\ell^N(\by|\theta)$ instead.
\end{itemize} 
In \cite{Koblents15} it was proved that, when the IWs can be computed exactly, the NIS approximation converges almost surely (a.s.) towards the target probability measure as $M \rw \infty$, which accounts for (i) above\footnote{The analysis of  \cite{Koblents15}  does not provide an error rate, though. Such rate is explicitly derived in this paper}. The problem of the approximate computation of the weights was partially addressed in \cite{Koblents16}, for a relatively simple case where the errors in the IWs where assumed deterministic and bounded. However, the estimation problem studied in \cite{Koblents16} (parameter estimation for $\alpha$-stable distributions using iid data) did not involve any dynamics and the convergence analysis only showed an upper bound for the approximation errors that included a deterministic constant, namely a non-vanishing term proportional to the approximation error of the IWs.

Here, we show stronger analytical results that ensure the almost sure convergence of the NIS approximation when $M\rw\infty$ and the likelihood function can only be estimated as $\ell^N(\by|\theta)$, i.e., using a BF with a finite and fixed number of particles $N$. Under assumptions which are standard in the classical IS theory, we prove that integrals of the form $\int f(\theta) \mu_{\by,k}^M(\sd\theta)$ converge towards $\int f(\theta) \mu_{\by,k}(\sd\theta)$ a.s. as $M\rw\infty$ and provide explicit error rates.

%% Notation
\subsection{Notation}

Since we focus our attention in the NIS scheme alone, i.e., a single iteration of the proposed algorithm, in the remaining of this section we drop the iteration index $k$. Hence, we assume a fixed importance density $q(\theta)$, from where $M$ independent  Monte Carlo samples, $\theta^1, \theta^2, \ldots, \theta^M$, are drawn. Since the observations $\by$ are assumed arbitrary but fixed, we drop them from the likelihood notation and write 
\begin{equation}
\ell(\theta)\dfn\ell(\by|\theta) \quad \mbox{and} \quad \ell^N(\theta)\dfn\ell^N(\by|\theta).
\end{equation} 
Similarly, we simplify the notation for the posterior pdf and write $p(\theta)=p(\theta|\by)$ and $\mu(\sd\theta)=\mu_\by(\sd\theta)$.
Then, the non-normalised IWs are approximated as 
\begin{equation}
\tilde \sw^i = g^N(\theta^i) \dfn \frac{\ell^N(\theta^i)p_0(\theta^i)}{q(\theta^i)},
\end{equation}
where we have introduced the weight function $g^N\dfn\ell^N p_0 / q$ as a shorthand. This weight function is a random approximation of the deterministic function $g=\ell p_0/q$. The support of $g$ is the same as the support of $q$, $\ell$ and $p_0$, denoted $\sS \subseteq \mathbb{R}^{d_\theta}$. We assume that $g(\theta) > 0$ for every $\theta \in \sS$ as well (a standard assumption in classical IS). It is also apparent that $p \propto gq$, where $p$ is the posterior pdf, and the proportionality constant is independent of $\theta$.

The non-normalised TIWs computed via the clipping function \eqref{eqClipping} are denoted 
\begin{equation}
\hat \sw^i = [\mT^M \circ g^N](\theta^i),
\end{equation}
where $\circ$ represents function composition and we omit the index argument of \eqref{eqClipping} for conciseness (its value is clear from the notation in any case). The normalised TIWs are $\sw^i = \frac{\hat \sw^i}{\sum_{j=1}^M \hat \sw^j}$, and they are used to compute the approximate measure $\mu^M(\sd\theta)=\sum_{i=1}^M \delta_{\theta^i}(\sd\theta) \sw^i$.
 
%% Assumptions
\subsection{Assumptions and a preliminary result}

Let the state sequence $\{\bx_n\}_{n\ge 0}$ take values on  $\mX \subseteq \Real^{d_x}$. We make the following classical assumptions on the conditional pdf of the observations $\by_n$, $n=1, 2, \ldots, R$, the prior density of the parameters, $p_0(\theta)$, and the importance function $q(\theta)$. 

%%%
\begin{hipotesis} \label{asBounds_on_g}
The observation sequence $\by_{1:R}$ is arbitrary but fixed. The functions $l_n(\by_n|\cdot):\mX\rw(0,\infty)$, $n=1, 2, ..., R$, are uniformly bounded, i.e., there exists a finite and positive constant $\| l \|_\infty$ such that
\begin{equation}
\| l \|_\infty = \sup_{n \ge 1, \bx_n \in \mX,\theta \in \sS} l_{n,\theta}(\by_n|\bx_n) < \infty.
\end{equation}
\end{hipotesis}
%%%
\begin{hipotesis} \label{asBounds_on_pq}
The ratio of pdf's $\frac{p_0(\theta)}{q(\theta)}$ is bounded on $\sS$, i.e., there exists a positive and finite constant $\left\| 
	\frac{p_0}{q}
\right\|_\infty$ such that
\begin{equation}
\left\| 
	\frac{p_0}{q}
\right\|_\infty = \sup_{\theta \in \sS} \left|
	\frac{
		p_0(\theta)
	}{
		q(\theta)
	}
\right| < \infty.
\end{equation}
\end{hipotesis}

%%%
\begin{nota}
If the parameter support set $\sS$ is compact, then A.\ref{asBounds_on_g} and A. \ref{asBounds_on_pq} hold naturally for most models of practical interest.
\end{nota}

The following lemma plays a key role in the asymptotic convergence analysis of the approximation $\mu^M(\sd\theta)$. It states that $\ell^N(\theta)$ is an unbiased estimator of the likelihood $\ell(\theta)$ and enables us to show that the NIS scheme converges when $M\rw\infty$, even if the number of particles $N$ in the approximation $\ell^N(\theta)$ remains finite and constant. 
%%%
\begin{lema} \label{lmUnbiased}
If Assumption \ref{asBounds_on_g} holds then 
\begin{equation}
\max \{ \ell(\theta), \ell^N(\theta) \} \le \| l \|_\infty^R < \infty \quad
\mbox{and} \quad
E\left[
	\ell^N(\theta)
\right] = \ell(\theta)
\end{equation}
independently of $N$. 
\end{lema}
\noindent\textbf{Proof.} From the definition of $\ell(\theta)$ in Eq. \eqref{eqDefEll} and its estimator $\ell^N(\theta)$ in Eq. \eqref{eqUBEstimator}, it is clear that both $\ell(\theta) \le \| l \|_\infty^R$ and $\ell^N(\theta) \le \| l \|_\infty^R$ when $R$ is the number of available observations. The fact that $\ell^N(\theta)$ is unbiased is a consequence of \cite[Theorem 7.4.2]{DelMoral04} (see also \cite[Lemma 2]{Crisan15par} for an alternative proof that does not rely on the Feynmann-Kac framework). \qed

%% Theorem
\subsection{Asymptotic convergence, error rates and exact approximation}

In the sequel we look into the approximation of integrals of the form 
\begin{equation}
(f,\mu) \dfn \int_\sS  f(\theta) \mu(\sd\theta),
\end{equation} 
where  $f$ is a bounded real function on the parameter space $\sS$. We use $\|f\|_\infty \dfn \sup_{\theta \in \sS}|f(\theta)| < \infty$ to denote the supremum norm of a bounded function, while the set of bounded functions on $\sS$ is denoted $B(\sS)$. The approximations of interest are
\begin{equation}
(f,\mu) \approx (f,\mu^M) = \sum_{i=1}^M f(\theta^i) \sw^i,
\end{equation}
for any $f \in B(\sS)$.

The following theorem yields an explicit upper bound for the (random) approximation error $| (f,\mu^M) - (f,\mu) |$. The bound is proportional to $M^{-\frac{1}{2}+\epsilon}$ (for an arbitrarily small $\epsilon>0$) and, therefore, it vanishes as $M \rw \infty$, independently of the number of particles $N$ used in the approximate likelihoods $\ell^N(\theta^i)$.

%%%
\begin{teorema} \label{thBasic}
Assume that A.\ref{asBounds_on_g} and A.\ref{asBounds_on_pq} hold, $M_c \le \sqrt{M}$ and $\int_\sS \ell(\theta)p_0(\theta)\sd\theta=(\ell,p_0)>0$. Then, for every $\epsilon \in \left(0,\frac{1}{2}\right)$ (arbitrarily small) and every $f \in B(\sS)$ there exists a positive and a.s. finite r.v. $V_{f,\epsilon}$, independent of $M$ and $M_c$, such that
\begin{equation}
| (f,\mu^M) - (f,\mu) | \le \frac{
    V_{f,\epsilon}
}{
    M^{\frac{1}{2}-\epsilon}
}.
\end{equation}
In particular, $\lim_{M\rw\infty} | (f,\mu^M) - (f,\mu) | =0$ a.s.
%\nonumber

\end{teorema}
%%%

%See Appendix \ref{apThBasic} for a proof.
\noindent\textbf{Proof.} Recall the intractable weight function $g=\ell p_0 / q$ and its random estimator $g^N = \ell^N p_0 / q$. The integral of any $f \in B(\sS)$ w.r.t. the posterior measure $\mu(\sd\theta) \propto \ell(\theta)p_0(\theta)\sd\theta$ can be written as
\begin{equation}
(f,\mu) = \frac{
    (fg,q)
}{
    (g,q)
}
\label{eq1}
\end{equation}
by simply noting that $g(\theta)q(\theta) = \ell(\theta)p_0(\theta)$. Similarly, for the random measure $\mu^M(\sd\theta)$ we can write
\begin{equation}
(f,\mu^M) = \frac{
    (f[\mT^M\circ g^N],q^M)
}{
    (\mT^M\circ g^N,q^M)
}
\label{eq2}
\end{equation}
where  $q^M(\sd\theta) = \frac{1}{M} \sum_{i=1}^M \delta_{\theta^i}(\sd\theta)$ is the Monte Carlo approximation of the proposal distribution (with pdf $q(\theta)$) and $\circ$ denotes composition of functions, hence $[\mT^M \circ g^N](\theta^i) = \mT^M( g^N (\theta^i) )$ is the transformed weight associated to $\theta^i$.

Given equations (\ref{eq1}) and (\ref{eq2}) it is straightforward to show that
\begin{equation}
(f,\mu^M) - (f,\mu) = \frac{
    (f[\mT^M \circ g^N],q^M) - (fg,q)
}{
    (g,q)
} + (f,\mu^M) \frac{
    (g,q) - (\mT^M \circ g^N,q^M)
}{
    (g,q)
}. \label{eq3}
\end{equation}
Since $(f,\mu^M) \le \| f \|_\infty<\infty$ and $(g,q)=(\ell,p_0)$, where $(\ell,p_0)  > 0$ by assumption, Eq. \eqref{eq3} readily yields
\begin{equation}
| (f,\mu^M) - (f,\mu) | \le \frac{
    1
}{
    (\ell,p_0)
} \left|
    (f[\mT^M \circ g^N],q^M) - (fg,q)
\right| + \frac{
    \| f \|_\infty
}{
    (\ell,p_0)
} \left|
    (\mT^M \circ g^N,q^M) - (g,q)
\right| \label{eq3_5}
\end{equation}
and, therefore, the problem of calculating bounds for $| (f,\mu^M) - (f,\mu) |$ reduces to the problem of computing bounds for errors of the form 
\begin{equation}
| (b[\mT^M \circ g^N],q^M)-(bg,q) |,
\end{equation}
 for $b \in B(\sS)$.

Choose any $b \in B(\sS)$. A simple triangle inequality yields
\begin{equation}
| (b[\mT^M\circ g^N],q^M) - (bg,q) | \le | (b[\mT^M\circ g^N],q^M) - (bg^N,q^M)  | + | (bg^N,q^M) - (bg,q) |. 
\label{eq4}
\end{equation}
It is straightforward to obtain an upper bound for the first term on the right hand side of the inequality (\ref{eq4}). Indeed, by construction of $\mT^M$ (see Eq. \eqref{eqClipping}) we readily obtain
\begin{eqnarray}
|(b[\mT^M\circ g^N], q^M) - (bg^N,q^M)| &=&
\left|
    \frac{1}{M}\sum_{r=1}^{M_c} b(\theta^{i_r}) \left[
        g^N(\theta^{i_{M_c}}) - g^N(\theta^{i_r})
    \right]
\right| \le \nonumber\\
&\le& 2 \| l \|_\infty^R \left\| \frac{p_0}{q} \right\|_\infty \|b\|_\infty \frac{M_c}{M}
\label{eq5}
\end{eqnarray}
where the inequality follows from the bound $g^N \leq \| l \|_\infty^R \left\| \frac{p_0}{q} \right\|_\infty$, which is a straightforward consequence of assumptions A.\ref{asBounds_on_g} and A.\ref{asBounds_on_pq} and the definition of the estimate $\ell^N$ produced by the BF (see Eq. \eqref{eqUBEstimator}).

Finding a suitable bound for the second term on the right hand side of the inequality (\ref{eq4}) takes some more effort. Choose, again, any $b \in B(\sS)$. A simple triangle inequality yields
\begin{equation}
| (bg^N,q^M) - (bg,q) | \le | (bg^N,q^M) - (bg,q^M) | + | (bg,q^M) - (bg,q) |. \label{eqTriang1}
\end{equation}
Since $q^M = \frac{1}{M} \sum_{i=1}^M \delta_{\theta^i}$, for the second term on the right hand side of \eqref{eqTriang1} we can write
\begin{equation}
\mbE\left[
    | (bg,q^M) - (bg,q) |^p
\right] = \mbE\left[
    \left|
        \frac{1}{M} \sum_{i=1}^M Z^i
    \right|^p
\right], \label{eqTriang2t}
\end{equation}
where the r.v.'s
\begin{equation}
Z^i = b(\theta^i)g(\theta^i) - (bg,q), \quad i=1, ..., M, \nonumber
\end{equation}
are independent, with zero mean (recall the $\theta^{(i)}$'s are i.i.d. draws from $q$) and bounded, because $b$ is bounded and A.\ref{asBounds_on_g} and A.\ref{asBounds_on_pq} imply that $g<\| l \|_\infty^R \times \left\|  \frac{p_0}{q} \right\|_\infty < \infty$. Therefore, it is an exercise in combinatorics to show that
\begin{equation}
\mbE\left[
    \left|
        \frac{1}{M} \sum_{i=1}^M Z^{(i)}
    \right|^p
\right] \le \frac{
    \tilde c^p \| l \|_\infty^{Rp} \left\|  \frac{p_0}{q} \right\|_\infty^p \| b \|_\infty^p
}{
    M^\frac{p}{2}
}, \label{eqZygmund}
\end{equation}
where $\tilde c$ is a constant independent of $M$ and $q$. Combining \eqref{eqZygmund} with \eqref{eqTriang2t} readily yields
\begin{equation}
\| (bg,q^M) - (bg,q) \|_p \le \frac{
    \tilde c \| l \|_\infty^R \left\|  \frac{p_0}{q} \right\|_\infty \| b \|_\infty
}{
    \sqrt{M}
}. \label{eq_tri2term}
\end{equation}
The inequality \eqref{eq_tri2term} implies that there exists an a.s. finite r.v. $\tilde U_{b,\epsilon}>0$ such that
\begin{equation}
| (bg,q^M) - (bg,q) | \le \frac{
    \tilde U_{b,\epsilon}
}{
    M^{\frac{1}{2}-\epsilon}
}, \label{eq_tri2term_eps}
\end{equation}
where $0 < \epsilon < \frac{1}{2}$ is an arbitrarily small constant independent of $M$ (see \cite[Lemma 4.1]{Crisan14a}).

If we expand the first term on the right hand side of \eqref{eqTriang1} we arrive at
\begin{eqnarray}
\left|
    (bg^N,q^M) - (bg,q^M)
\right| &=& \left|
    \frac{1}{M} \sum_{i=1}^M b(\theta^i) \left(
    		g^N(\theta^i) - g(\theta^i)
	\right)
\right| \nonumber \\
&=& \left|
    \frac{1}{M} \sum_{i=1}^M Z_N^i
\right|, \label{eqZ2}
\end{eqnarray}
where the r.v.'s $Z_N^i = \frac{b(\theta^i)p_0(\theta^i)}{q(\theta^i)}\left(  \ell^N(\theta^i) - \ell(\theta^i) \right)$, $i=1, 2, ..., M$, are independent (because the samples $\theta^1, \ldots, \theta^M$ are independent) and zero mean, as a result of Lemma \ref{lmUnbiased}\footnote{Note that 
$E\left[Z_N^i|\theta^i\right]=\frac{b(\theta^i)p_0(\theta^i)}{q(\theta^i)}E\left[ \ell^N(\theta^i)- \ell(\theta^i) \right]=0$, because $\ell^N(\theta^i)$ is an unbiased estimator of $\ell(\theta^i)$, hence $E\left[Z_N^i \right]=E\left[ E\left[Z_N^i|\theta^i \right] \right]=0$.}. Since they are also bounded, namely $| Z_N^i | \le \| b \|_\infty \| l \|_\infty^R \left\| \frac{p_0}{q}\right\|_\infty$ as a consequence of A.\ref{asBounds_on_g} and A.\ref{asBounds_on_pq}, it is again an exercise to show that \eqref{eqZ2} implies
\begin{equation}
E\left[
	\left|
    		(bg^N,q^M) - (bg,q^M)
	\right|^p
\right] \le \frac{
    \bar c^p \| l \|_\infty^{Rp} \left\|  \frac{p_0}{q} \right\|_\infty^p \| b \|_\infty^p
}{
    M^\frac{p}{2}
}
\label{eqZygmund2}
\end{equation}
in the same manner as we obtained the inequality \eqref{eqZygmund}. Resorting again to \cite[Lemma 4.1]{Crisan14a}, from \eqref{eqZygmund2} we deduce that there exists an a.s. finite r.v. $\bar U_{b,\epsilon}>0$, independent of $M$, such that
\begin{equation}
| (bg^N,q^M) - (bg,q^M) | \le \frac{
    \bar U_{b,\epsilon}
}{
    M^{\frac{1}{2}-\epsilon}
}, \label{eq_tri1term_eps}
\end{equation}
where $0 < \epsilon < \frac{1}{2}$ is an arbitrarily small constant independent of $M$.
 
Taking together \eqref{eqTriang1}, \eqref{eq_tri2term_eps} and \eqref{eq_tri1term_eps} we arrive at
\begin{equation}
| (bg^N,q^M) - (bg,q) | \le \frac{
    U_{b,\epsilon}
}{
    M^{\frac{1}{2}-\epsilon}
}, \label{eq6}
\end{equation}
where $U_{b,\epsilon}=\tilde U_{b,\epsilon} + \bar U_{b,\epsilon} \ge 0$ is an a.s. finite r.v. independent of $M$, and $\epsilon \in \left(0,\frac{1}{2}\right)$ can be chosen to be arbitrarily small.   

Substituting the inequalities (\ref{eq5}) and (\ref{eq6}) back into the relation (\ref{eq4}) we arrive at the bound
\begin{equation}
|(b[\mT^M\circ g^N], q^M) - (bg,q)|  \le 2 \| l \|_\infty^R \left\| \frac{p_0}{q} \right\|_\infty \|b\|_\infty \frac{M_c}{M}+\frac{
    U_{b,\epsilon}
}{
    M^{\frac{1}{2}-\epsilon}
}\le  \frac{
    \tilde V_{b,\epsilon}
}{
    M^{\frac{1}{2}-\epsilon}
}
\label{eq7}
\end{equation}
where the second inequality follows from the assumption $M_c\le\sqrt{M}$ and choosing $\tilde V_{b,\epsilon}=2 \| l \|_\infty^R \left\| \frac{p_0}{q} \right\|_\infty \|b\|_\infty + U_{b,\epsilon}$. Since the r.v. $U_{b,\epsilon}$ is a a.s. finite, $\tilde V_{b,\epsilon}<\infty$ a.s. as well.

To conclude the proof, we substitute the inequality (\ref{eq7}) twice into the relation (\ref{eq3_5}). To be precise, we choose $b=f$ first and use (\ref{eq7}) to obtain a bound for the first term on the right hand side of (\ref{eq3_5}). Then, we choose $b=1$ and apply  (\ref{eq7}) again to find a bound for the second term on the right hand side of (\ref{eq3_5}). As a result, we arrive at
\begin{equation}
| (f,\mu^M) - (f,\mu) | \le \frac{
    \tilde V_{f,\epsilon}
}{
    (\ell,p_0)
}
\times
\frac{
    1
}{
    M^{\frac{1}{2}-\epsilon}
}
+
 \frac{
    \| f \|_\infty \tilde V_{1,\epsilon}
}{
    (\ell,p_0)
}
\times
\frac{
    1
}{
    M^{\frac{1}{2}-\epsilon}
}.
\end{equation}
Since $(\ell,p_0)>0$ by assumption of Theorem \ref{thBasic}, taking 
\begin{equation}
 V_{f,\epsilon}=\frac{
    1
}{
    (\ell,p_0)
}
\left( 
\tilde V_{f,\epsilon}+ \| f \|_\infty \tilde V_{1,\epsilon}\right)<\infty \quad \mbox{a.s.}
\end{equation}
 leads to the desired result and concludes the proof. \qed

%\begin{nota}
Theorem \ref{thBasic} is a general result regarding nonlinear importance sampling. It holds true for any problem involving the approximation of the posterior probability distribution of the unknown parameters of a state space model as long as Assumptions 1 and 2 hold. These assumptions, in turn, are very mild and amount to the classical assumptions in the analysis of standard IS algorithms. 
%\end{nota}

\begin{nota}
We draw attention to the fact that the error $| (f,\mu^M) - (f,\mu) |$ vanishes a.s. when $M\rightarrow\infty$ even if the number of particles $N$ in the BF remains fixed and, hence, $\ell^N$ does not converge to $\ell$. This property has been coined ``exact approximation" in the MCMC literature (see \cite{Andrieu10}).
\end{nota}
%%%%%%

%%%%%

%%%%%
%\section{Computer simulations} \label{sSimulations}

\section{Computer simulations}
\label{sec:simulations}

\subsection{State-space models} \label{ssmodels}
In order to illustrate the performance of the NPMC algorithm and the exact approximation property granted by Theorem \ref{thBasic} we have carried out computer simulations for the estimation of the unknown parameters in a problem consisting of the tracking of a target moving over a region monitored by a network of sensors.
%of the signal transmitted by the target.

\subsubsection{Target dynamics} \label{starget}
The target moves over a closed rectangular region ${\mathcal R} = [-20,+20] \times [-10,+10]$. When it hits the border of ${\mathcal R}$, the target bounces back in according to the law of reflection \cite{Farin13}. The state of the system at time $n$ is 
$\bx_n=\left[
	\begin{array}{c}
	{\bf r}_n\\ 
	{\bf v}_n\\
	\end{array}
\right] \in \Real^4,$
where ${\bf r}_n \in {\mathcal R}$ is the target position and ${\bf v}_n$ its velocity. At time $n=0$, we assume a uniform prior on  ${\mathcal R}$ for the position and a zero-mean Gaussian distribution for the velocity. To be specific, the prior probability measure is defined as
\begin{equation}
%	\sfk_0(dx)
	{\mathcal K}_0({\sf d}\bx_0)
	=
	{\mathcal U}({\mathcal R}) \times {\mathcal N}({\bf 0},\frac{1}{20}\times{\bf I}_2)
	\label{eq:priorSim}
\end{equation}
where ${\bf I}_2$ is the $2\times2$ identity matrix, ${\mathcal U}({\mathcal R})$ is the uniform distribution on ${\mathcal R}$ and ${\mathcal N}({\bf m},{\bf C})$ denotes the Gaussian distribution with mean $\bf m$ and covariance matrix  $\bf C$. 

At time $n>0$, the state vector $\bx_n$ evolves according to a linear-Gaussian equation if the target position remains within the bounded region ${\mathcal R}$ but it ``reflects" back in when the target reaches a border of ${\mathcal R}$. Specifically, let
\begin{equation}
\tilde \bx_n=\left[
	\begin{array}{c c}
	{\bf I}_2 & \kappa{\bf I}_2\\
	0 & {\bf I}_2\\
 	\end{array}
	\right]\bx_{n-1}+{\bf u}_n,
	\label{eq:transitionKernel}
\end{equation}
where ${\bf u}_n \sim {\mathcal N}({\bf 0},{\bf C})$ is a Gaussian noise term with $0$-mean and covariance matrix
\begin{equation}
{\bf C}=\left[
	\begin{array}{c c}
	(\kappa\sigma_u^2+\sigma_z^2){\bf I}_2 & 0\\
	0 & \sigma_u^2{\bf I}_2\\
 	\end{array}
	\right],
\end{equation}
$\kappa$ is a time-discretisation step (we assume $\kappa=1$ in our simulations), $\sigma_u^2$ is a velocity variance parameter, and $\sigma_z^2$ is a position variance parameter. The latter are assumed known and identical, $\sigma_u^2=\sigma_z^2=10^{-2}$. If $\tilde \bx_n$ generated in this way is inside 
${\mathcal R}$, $\tilde \bx_n \in{\mathcal R}$, then $\bx_n=\tilde \bx_n$, otherwise $\bx_n=f(\bx_{n-1})$, where $f$ is the reflection function detailed in \ref{appendix:fuctionDef}. Note that we do not provide an expression for the kernel ${\mathcal K}_n({\sf d}\bx_n|\bx_{n-1})$ but have just described how to draw samples from it instead. This is enough for the implementation of the bootstrap filter and the NPMC algorithm.

\begin{figure}[htpb]
	\pgfplotstableread{TR_trajectory.txt}\trajectory
	\pgfplotstableread{TR_sensors.txt}\sensorsPositions
	\begin{center}
		\begin{tikzpicture}
		
			\colorlet{trajectory}{brown!20!black}
		
			\begin{axis}[
%			    x=4,y=4,
			    xmin=-20,
			    xmax=20,
			    ymin=-10,
			    ymax=10,
		%	    view={0}{90},
			    axis background/.style={fill=white}
			]
			
			\addplot[decoration={
			    markings,
			    mark=between positions 0.1 and 1 step 3em with {\arrow [scale=1.5]{stealth}}
			    }, postaction=decorate, color=trajectory] table from \trajectory;
			    
			\addplot[only marks,mark = square*,mark options={blue}] table from \sensorsPositions;
			   
			% initial position is extracted from the file
			\pgfplotstablegetelem{0}{[index]0}\of{\trajectory}
			\pgfmathsetmacro{\startX}{\pgfplotsretval}
			\pgfplotstablegetelem{0}{[index]1}\of{\trajectory}
			\pgfmathsetmacro{\startY}{\pgfplotsretval}
			
			\node[draw, shape=diamond,fill=red,above,inner sep=0,outer sep=0,minimum size=4] at (axis cs: \startX,\startY) {};
			   
			\end{axis}

		\end{tikzpicture}
		\caption{
		Wireless sensors network with a sample trajectory overimposed. 
		The blue squares mark the positions of the sensors, and the red diamond indicates the starting point of the trajectory, which is depicted as a black solid line.
		}
		\label{fig:trajectory}
	\end{center}
\end{figure}
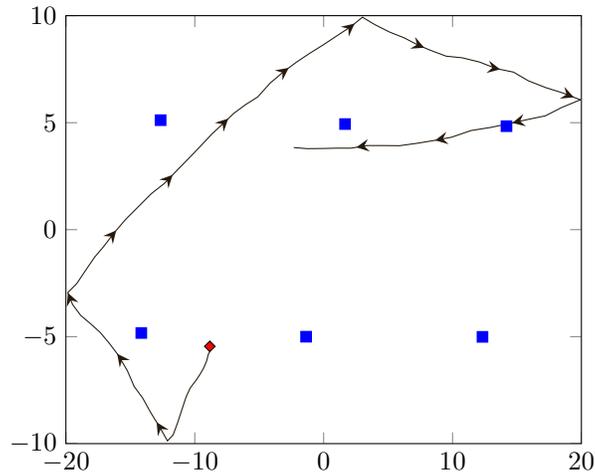

For illustration, Fig. \ref{fig:trajectory} depicts the region ${\mathcal R}$ and a sample trajectory (i.e., a sequence of positions ${\bf r}_0, {\bf r}_1, \ldots$) which hits the borders of  ${\mathcal R}$ and is reflected back in at four different times. In the figure, the starting target position is represented by a red diamond, the direction of motion is indicated by arrows and the blue squares represent the position of the sensors used to monitor the target motion.

%%%%% FIGURE 1 (TRAJECTORY)

\subsubsection{Observations}\label{sobservations}
There are $J$ sensors deployed in  ${\mathcal R}$ and, at time $n$, each sensor collects a measurement of the power of the radio signal transmitted by the target. To be specific, the observation recorded by sensor $j$ at time $n$ has the form
\begin{equation}
y_{j,n}=10\log\left (\frac{P_t}{||{\bf r}_n-{\bf s}_j||^\nu}+\rho \right )+\epsilon_{j,n}
\end{equation}
where $P_t$ is the power of the transmitted radio signal, ${\bf s}_j$ is the location of the $j$th sensor, $||{\bf r}_n-{\bf s}_j||$ is the distance at time $n$ between the target and the sensor, $\nu>0$ is the path loss exponent, $\rho$ is the sensitivity of the sensor, i.e., the minimum power it can measure (note that $y_{j,n}\rw 10\log(\rho)+\epsilon_{j,n}$ when $||{\bf r}_n-{\bf s}_j|| \rw\infty$) and $\epsilon_{j,n}\sim{\mathcal N}(0,\sigma_\epsilon^2)$ is a Gaussian term accounting for observational errors. We assume $\sigma_\epsilon^2=1$ is a known parameter.

At each time instant $n$, a vector of $J$ observations ${\bf y}_n=[y_{1,n},y_{2,n},\ldots,y_{J,n}]^T\in\Real^J$ is collected. The target is observed over $m$ time instants, and hence the available dataset is ${\bf y}={\bf y}_{1:m}$. We set $m=50$ for our computer simulations.
%%%%%%
%
%%%%%%
\subsubsection{Problem statement}

Given the state space model described in Sections  \ref{starget} and \ref{sobservations} above, we aim at estimating the unknown parameters $P_t$, $\nu$ and $\rho$. All other parameters (namely the discretisation period $\kappa$ and the relevant variances) are assumed known. For all computer simulations we have set ground truth  values $P_t=0.8$, $\nu=3$ and $\rho=10^{-5}$ for the parameters to be estimated.

Since $P_t>0$ and $\rho>0$, we apply the NPMC algorithm (together with competing algorithms to be described below) to approximate the posterior probability measure $\mu_{\by}(\sd \theta)$ of the vector of unknowns $\theta=[\log P_t,\nu,\log \rho]^T\in\Real^3$. We assume prior distributions of the form $\log P_t\sim{\mathcal N}(-0.11,0.22)$, $\nu\sim{\mathcal N}(0,4)$ and $\log \rho \sim{\mathcal N}(-11.02,0.4)$. Note that, in natural units, the prior mean and variance of $P_t$ are $1$ and $0.25$, respectively, while for $\rho$ the prior mean and variance are $2\times10^{-5}$ and $2\times 10^{-10}$.

The likelihood $\ell(\by|\theta)$ for the model does not have a closed form and, therefore, it is estimated using a BF, for the state space model described in Sections \ref{starget} and \ref{sobservations}, to yield the approximation  $\ell^N(\by|\theta)$ detailed in Section  \ref{ssFilter}.

%%%%%%
%
%%%%%%
\subsection{Competing methods}

We have applied to this problem the NPMC method described in Section \ref{sNPMC}, a standard PMC procedure and a particle Metropolis-Hastings (pMH) algorithm. The PMC scheme we have used is identical to the NPMC algorithm of Section   \ref{sNPMC} except that TIWs are not computed, hence all approximations rely on the conventional IWs.

The pMH is a representative of the class of particle MCMC methods \cite{Andrieu10} that have become popular in the past two years. It generates a Markov chain on the space of the unknown parameter vector $\theta$ according to the following procedure:
\begin{enumerate}
\item Draw $\theta_0\sim p_0(\theta)$ from the prior distribution of the parameters
\item At the $r$-th iteration, and given the previous element $\theta_{r-1}$:
\begin{enumerate}
\item Draw a tentative new element $\tilde	\theta_r\sim{\mathcal N}(\theta_{r-1},\frac{2}{10}\bf C)$, where both
$
\bf C
=
%[\ldots]
\diag{\left[0.22,4,0.4\right]}
%\begin{bmatrix}
%0.22 & 0 & 0
%\\
%0 & 4 & 0
%\\
%0 & 0 & 0.4
%\end{bmatrix}
$
and the scale factor $\frac{2}{10}$ have been empirically chosen to optimise the performance of the algorithm. 
\item Compute the (approximate) likelihood $\ell^N(\by|\tilde\theta_r)$ and prior density $p_0(\tilde\theta_r)$. The acceptance probability for $\tilde\theta_r$ is 
\begin{equation}
\alpha_r=\min\left ( 1, \frac{\ell^N(\by|\tilde\theta_r)p_0(\tilde\theta_r)}{\ell^N(\by|\theta_{r-1})p_0(\theta_{r-1})}\right )
\end{equation}
\item Draw $u_r\sim {\mathcal U}(0,1)$. If $u_r<\alpha_r$ then $\theta_r=\tilde\theta_r$, else $\theta_r=\theta_{r-1}$. 
\end{enumerate}
\end{enumerate}
When we generate a chain of length $L$ using the procedure above we set a burn-in period of $L\over2$, hence estimates are computed from the samples $\theta_{\lfloor {L\over2} \rfloor +1},\ldots,\theta_L$ in the chain. 

To compare the pMH and PMC-like algorithms on a fair basis, we let $L=M\times K$, where $K$ is the number of iterations of the NPMC and PMC algorithms and $M$ is the number of samples generated per iteration.

All three methods (PMC, NPMC, pMH) rely on a BF with $N$ particles for the computation of $\ell^N({\bf y}|\theta)$. The value of $N$ is fixed for all algorithms as $N=400$ unless explicitly stated otherwise.

%%%%%%%%%%%%%
\subsection{Results}

\begin{figure}[htpb]
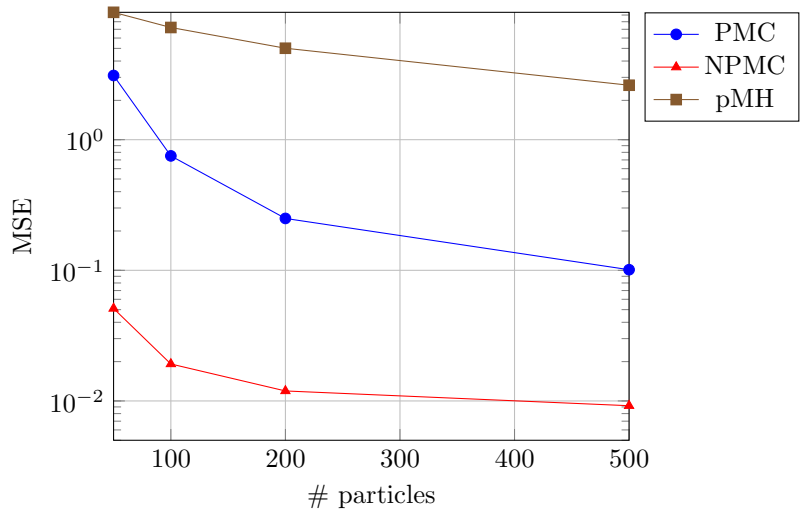

	\begin{center}
		\tikz \pic{vs={mse_vs_particles.txt}{particles}{particles,mse,ymin=0.005}};
		\caption{
			MSE for several values of $M$. The PMC and NPMC algorithms are iterated $K=10$ times. The pMH scheme generates a chain of length $L=M\times K$. The curves are averaged over 1,000 independent simulation runs. 
		}
		\label{fig:mse_vs_samples}
	\end{center}
\end{figure}

Figure \ref{fig:mse_vs_samples} shows the evolution of the MSE of the estimators of $\theta$ produced by the PMC, NPMC and pMH algorithms as the number of samples is increased.

%%%% FIGURE 2: MSE VS M

The error for the NPMC algorithm is at least one order of magnitude below the errors of the conventional PMC and the pMH algorithms for every tested value of $M$. For $M=200$ samples, for example, the MSE attained by the NPMC is $\approx 1.19\times 10^{-2}$, while for the standard PMC and pMH algorithms the errors are $\approx 2.49\times 10^{-1}$ and $\approx 5.01$, respectively.

\begin{figure}[htpb]
	\begin{center}
		\begin{tikzpicture}
			\pgfplotstableread{vsMH_mse_vs_particles.txt}\mhTable
			
			\pgfplotstableread{vsNPMC_mse_vs_particles.txt}\npmcTable
			
			% the element it the 1st row and 2nd column (numbering starts at 0): the MSE for the NPMC
			\pgfplotstablegetelem{0}{[index]1}\of{\npmcTable}
			\pgfmathsetmacro{\npmcThresh}{\pgfplotsretval}
			
			% the number of "samples" for the NPMC (it is stored in \pgfplotsretval after this line)
			\pgfplotstablegetelem{0}{[index]0}\of{\npmcTable}
			
			% algorithm NPMC uses 10 iterations
			\pgfmathparse{\pgfplotsretval*10}\pgfmathresult
			\pgfmathsetmacro{\npmcSamples}{\pgfmathresult}
			
			% vertical space between lines
			\linespread{1.1}
			
			\begin{loglogaxis}[
				ymin=0.002,
				legend pos=outer north east,
				enlargelimits=false,
				grid=major,
				log ticks with fixed point,
				particles,
%				bias,
				mse
				]
				
				% curve for the MH is added
				\addplot[color=MH,mark=\MHmark] table[x=particles,y=MetropolisHastings] from \mhTable;
				\addlegendentry{pMH};
				
				% horizontal line for NPMC
				\addplot[dashed, mark=none,domain=50:1000000] {\npmcThresh};
				
				% vertical line
				\draw[dashed] ({axis cs:\npmcSamples,0}|-{rel axis cs:0,0}) -- ({axis cs:\npmcSamples,0}|-{rel axis cs:0,1});
				
				% coordinate and node for the marked NPMC point
				\coordinate (npmc) at (axis cs:\npmcSamples, \npmcThresh);
				\node[shape=circle,fill=NPMC,minimum size=3,inner sep=0,outer sep=0] at (npmc) (npmc point) {};
				
				% label for the point
				\node[above left=0.75 and 0.3 of npmc, align=center,color=NPMC] (label) {NPMC\\$\nParticles=500$, $\nIterations=10$};
				
				% arrow between the label and the point
				\draw[->] (label.south) to[bend left=15,gray] (npmc point);
				
			\end{loglogaxis}
		\end{tikzpicture}
		\caption{
			MSE for different numbers of chain lengths, $L$, of the pMH algorithm. These results have been averaged over 100 independent simulation runs.
		}
		\label{fig:bias_vs_samples_mh}
	\end{center}
\end{figure}
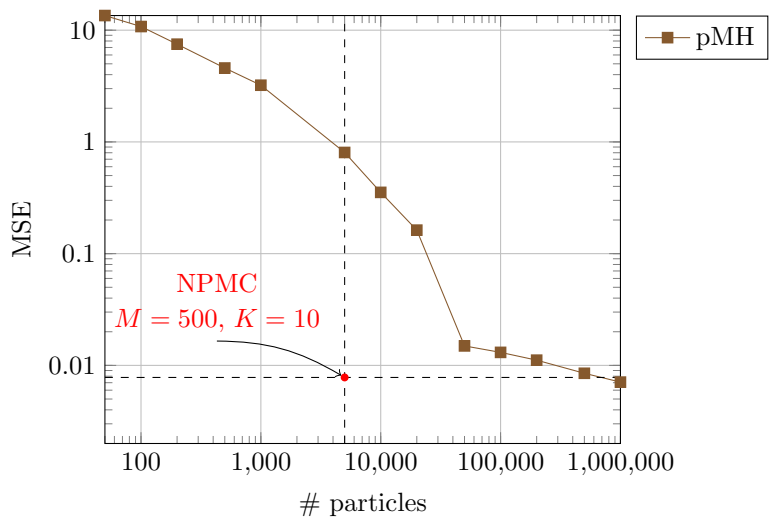

%In the next experiment we assess the number of samples required by the pMH algorithm to attain the same performance as the NPMC.
%The results shown so far suggest the performance of the Metropolis algorithm 
%would improve if longer chains were to be used. In the last experiment, we aim at finding out the number of samples required for this algorithm to reach the same performance, in terms of bias, as the NPMC algorithm.
Next, we aim at finding out the length of the chain, $L$, required for the pMH algorithm to attain the same performance, in terms of MSE, as the NPMC algorithm.  Figure \ref{fig:bias_vs_samples_mh} shows the MSE of the pMH method for different chain lengths (equivalently, number of generated samples).

%%%%% FIGURE 3: MSE VS CHAIN LENGTH

For comparison, the performance of the NPMC algorithm for $M=500$ samples and $K=10$ iterations ($500\times 10 = 5,000$ Monte Carlo samples overall) is also indicated in the plot. 
%It can be seen that the Metropolis-Hastings algorithm requires chains that are around $500,000$ samples long to attain the same MSE as the NPMC algorithm, .
It can be seen that, in the pMH algorithm, chains that are around $500,000$ samples long are required to attain the same MSE as the NPMC algorithm (a 100-fold increase of the computational cost). While the parameters of the pMH scheme may be further tuned to improve this performance, the gap between the algorithms is large enough to conclude that the NMPC method is more efficient in this example.

\begin{figure}[htpb]
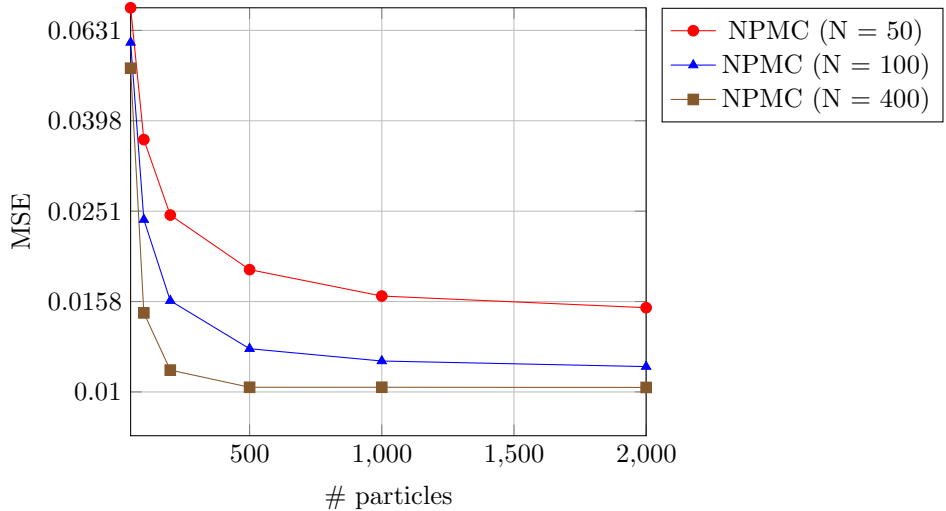

	\begin{center}
		\tikz \pic{NPMC={NPMC_mse_vs_particles.txt}{particles}{particles,mse,ymin=0.008}};
		\caption{
			MSE vs. number of samples, $M$, attained by the NPMC algorithm with different choices of the number of particles in the BF, $N$. The curves are averaged over 100 independent simulation runs.
		}
		\label{fig:NPMC_mse_vs_samples}
	\end{center}
\end{figure}

Finally, we examine the exact approximation property of the NPMC scheme stated by Theorem \ref{thBasic}. Figure \ref{fig:NPMC_mse_vs_samples} shows the MSE of the NPMC algorithm versus the number of Monte Carlo samples, $M$, for different values of $N$ (the number of particles used by the BF to approximate the IWs). While Theorem \ref{thBasic} guarantees that the approximation errors vanish as $M\rw\infty$, even if $N$ is fixed, it is reasonable to expect that for a fixed $M<\infty$, greater values of $N$ lead to better performance. This is shown, indeed, by Fig. \ref{fig:NPMC_mse_vs_samples}. Note, however, that the difference in performance is very small. For $M=1,000$, the gap between the MSE of the NPMC scheme with $N=400$ and the NPMC scheme with $N=50$ is $\approx 6 \times 10^{-3}$.

%%%%% FIGURE 4: MSE VS M, FOR SEVERAL CHOICES OF N

%%%%%%%%%%%%

\section{Conclusion} \label{sConclusions}

We have rigorously proved, under mild assumptions,  that nonlinear importance samplers with clipped IWs  converge a.s. with   optimal Monte Carlo error rates even when the  weights can only be estimated (and have a positive, non-vanishing variance) as long as these estimates are unbiased. Therefore, nonlinear importance samplers can perform exact approximation in the same manner as, e.g., particle MCMC schemes. Besides the theoretical contribution, we have numerically shown that the proposed algorithm can be more efficient than a particle Metropolis-Hastings algorithm of the same complexity for inference on a target tracking model. 

\section*{Acknowledgments}
This research has been partially supported by the Spanish Ministry of Economy and Competitiveness (projects TEC2015- 69868-C2-1-R ADVENTURE and FIS2013-40653-P), the Spanish Ministry of Education, Culture and Sport (mobility award PRX15/00378) and the Office of Naval Research (ONR) Global (Grant Award no. N62909-15-1-2011). %, the regional Government of Madrid (program CASI-CAM-CM S2013/ICE-2845).

\begin{appendices}

\section{Definition of function $\transitionKernel(\cdot)$}
\label{appendix:fuctionDef}

Let us denote the upper right, upper left, lower left and lower right vertices of the monitored region by, respectively, $\bc_0$, $\bc_1$, $\bc_2$ and $\bc_3$. The sides of the rectangle, obtained by joining adjacent vertices, are denoted $\bl_0 =\overline{\bc_1\bc_0}$ (top), $\bl_1=\overline{\bc_1\bc_2}$ (left), $\bl_2=\overline{\bc_2\bc_3}$ (bottom) and $\bl_3=\overline{\bc_3\bc_0}$ (right). With this notation, Algorithm \ref{alF} can be used at time $n$ to generate a sample $\bx_n=[\br_n^\top,\bv_n^\top]^\top$ from $\bx_{n-1}=[\br_{n-1}^\top, \bv_{n-1}^\top]^\top$. It accounts for the scenario in which the target hits one of the walls and deals with it by means of the law of reflection \cite{Farin13}.

\begin{algorithm}
	\caption{
		Generation of a sample $\bx_n \in {\mathcal R}$, conditional on $\bx_{n-1}$
	}
	\label{alF}
	\begin{algorithmic}[1]
		\STATE Generate $\tilde \bx_n = [\tilde \br_n^\top, \tilde \bv_n^\top]^\top$, conditional on $\bx_{n-1}$, using Eq. \eqref{eq:transitionKernel}.
		\STATE If $\tilde {\bf r}_n \in {\mathcal R}$ then return $\bx_n = f(\bx_{n-1}) = \tilde \bx_n$. Otherwise, continue.
		\STATE Compute the vectors
		\[
			\begin{array}{ccc}
				\bq_j
				=
				\bc_j -
				\br_{n-1},
				\quad j=0, 1, 2, 3,
				&
				\text{and}
				&
				\bs
				=
				\tilde \br_n  - 
				\br_{n-1}
			\end{array}
		\]
		and the corresponding angles
		\[
			\Theta_s
			=
			\angle(\bs),
			\quad \Theta_j = \angle(\bq_j), \quad j=0,1,2,3,
		\]
		i.e., the angles of vectors $\bs$ and $\bq_j$, respectively, w.r.t. the horizontal axis
		\label{algo:computeAngles}
		\STATE Find $j \in \{0,1,2,3\}$ such that $\Theta_j < \Theta_s < \Theta _{(j+1) \text{{\sf mod} } 4}$ and decompose $\tilde \br_n$ as
		\[
			\tilde \br_n = \br_{n-1} + \bs' + \bs'',
		\]
		where $\bs' = \Lambda(j)\bs$, $s'' = (1-\Lambda(j))\bs$ and
		\[
		\Lambda(j) = \left\{
			\begin{array}{ll}
			(\bc_j(2) - \br_{n-1}(2))/\bs(2), &\text{for } j=0, 2\\
			(\bc_j(1) - \br_{n-1}(1))/\bs(1), &\text{for } j=1,3\\
			\end{array}
		\right.
		\]
		(with $b(j)$ denoting the $j$-th component of vector $b$). 
		\label{algo:findQuadrant}
		\STATE Compute the vector ${\sf n}_j$ normal to $\bl_j$ (namely ${\sf n}_j^\top \bl_j = 0$ and $\| {\sf n}_j \| = 1$).
		Compute the new state vector $\bx_n=[\breve \br_n^\top, \breve \bv_n^\top]^\top$, where
		\begin{align*}
			\begin{array}{cc}
				\breve \br_n
				=
				\br_{n-1} + \bs' + \bs'' - 2{\sf n}_j{\sf n}_j^\top \bs'',
				&
				\breve \bv_n
				=
				\frac{
					\bs'' - 2{\sf n}_j{\sf n}_j^\top \bs''
				}{
					\| \bs'' - 2{\sf n}_j{\sf n}_j^\top \bs'' \|
				} \| \tilde \bv_n \|
			\end{array}
		\end{align*}
		\label{algo:computeNormalVector}
		\RETURN $f(\bx_{n-1})=\bx_n$
	\end{algorithmic}
\end{algorithm}

We are implicitly  assuming that $\br_n \in {\mathcal R}$ in step 5 above. If this is not the case, i.e., $r_n \notin {\mathcal R}$, then steps 3--5 can be run again to implement a second reflection.

\end{appendices}

%-------------------------------------
\section*{References}

  \bibliographystyle{elsarticle-num}
% \bibliography{bibliography}
\bibliography{./bibliografia}

%%%%%%%%%% FIGURES

% FIG 1 (TRAJECTORY)

% FIG. 2: MSE VS M

% FIG. 3: MSE VS CHAIN LENGTH

% FIG. 4: MSE VS M FOR VARIOUS N's

\end{document}

%% file: manu_preamble.tex
\usepackage{pgfplots}
\pgfplotsset{compat=1.3}
\input{plots_defs}
\pgfplotsset{
}

\usetikzlibrary{decorations.markings,shapes,backgrounds,positioning}

\newcommand*{\diag}[1]{\text{diag}\left(#1\right)}

\newcommand{\transitionKernel}{f}

\newcommand{\nParticles}{M}
\newcommand{\nIterations}{K}

\newcommand{\br}{{\bf r}}
\newcommand{\bv}{{\bf v}}
\newcommand{\bq}{{\bf q}}
\newcommand{\bc}{{\bf c}}
\newcommand{\bs}{{\bf s}}
\newcommand{\bl}{{\bf l}}

%% file: plots_defs.tex
\colorlet{PMC}{blue}
\colorlet{NPMC}{red}
\colorlet{MH}{brown!70!black}

\newcommand{\PMCmark}{*}
\newcommand{\NPMCmark}{triangle*}
\newcommand{\MHmark}{square*}

\colorlet{NPMCFifty}{red}
\colorlet{NPMCHundred}{blue}
\colorlet{NPMCFourHundred}{brown!70!black}

\newcommand{\NPMCmarkFifty}{*}
\newcommand{\NPMCmarkHundred}{triangle*}
\newcommand{\NPMCmarkFourHundred}{square*}

\tikzset
{
	pics/vs/.style
		n args={3}{
			code={
				\pgfplotstableread{#1}\loadedtable
				\begin{semilogyaxis}[
					legend pos=outer north east,
					enlargelimits=false,
					grid=major,
					ymin=0.05,
					#3
					]
					
					\addplot[color=PMC,mark=\PMCmark] table[x=#2,y=PMC] from \loadedtable;
					\addlegendentry{PMC}
				
					\addplot[color=NPMC,mark=\NPMCmark] table[x=#2,y=NPMC] from \loadedtable;
					\addlegendentry{NPMC}

					\addplot[color=MH,mark=\MHmark] table[x=#2,y=MetropolisHastings] from \loadedtable;
					\addlegendentry{pMH}
					
				\end{semilogyaxis}
		}
	},
	pics/vs only PMC/.style
		n args={3}{
			code={
				\pgfplotstableread{#1}\loadedtable
				\begin{semilogyaxis}[
					legend pos=outer north east,
					enlargelimits=false,
					grid=major,
					log ticks with fixed point,
					#3
					]
					
					\addplot[color=PMC,mark=\PMCmark] table[x=#2,y=PMC] from \loadedtable;
					\addlegendentry{PMC}
				
					\addplot[color=NPMC,mark=\NPMCmark] table[x=#2,y=NPMC] from \loadedtable;
					\addlegendentry{NPMC}
					
				\end{semilogyaxis}
		}
	},
	pics/NPMC/.style
		n args={3}{
			code={
				\pgfplotstableread{#1}\loadedtable
				\begin{semilogyaxis}[
					legend pos=outer north east,
					enlargelimits=false,
					grid=major,
					ymin=0.01,
					log ticks with fixed point,
					#3
					]
					
					\addplot[color=NPMCFifty,mark=\NPMCmarkFifty] table[x=#2,y=NPMC_50] from \loadedtable;
					\addlegendentry{NPMC (N = 50)}
					
					\addplot[color=NPMCHundred,mark=\NPMCmarkHundred] table[x=#2,y=NPMC_100] from \loadedtable;
					\addlegendentry{NPMC (N = 100)}
					
					\addplot[color=NPMCFourHundred,mark=\NPMCmarkFourHundred] table[x=#2,y=NPMC_400] from \loadedtable;
					\addlegendentry{NPMC (N = 400)}
					
				\end{semilogyaxis}
		}
	},
}

\pgfplotsset{
	iterations/.style={
		xlabel=\# iterations
	},
	bias/.style={
			ylabel=bias
	},
	particles/.style={
		xlabel=\# particles
	},
	variance/.style={
			ylabel=variance
	},
	max weight/.style={
			ylabel=maximum weight
	},
	effective sample size/.style={
			ylabel=effective sample size
	},
	mse/.style={
			ylabel=MSE
	}
}

%% file: ines.bbl
\begin{thebibliography}{10}
\expandafter\ifx\csname url\endcsname\relax
  \def\url#1{\texttt{#1}}\fi
\expandafter\ifx\csname urlprefix\endcsname\relax\def\urlprefix{URL }\fi
\expandafter\ifx\csname href\endcsname\relax
  \def\href#1#2{#2} \def\path#1{#1}\fi

\bibitem{Jansson96}
M.~Jansson, B.~Wahlberg, A linear regression approach to state-space subspace
  system identification, Signal Processing 52~(2) (1996) 103--129.

\bibitem{Storvik02}
G.~Storvik, Particle filters for state-space models with the presence of
  unknown static parameters, IEEE Transactions Signal Processing 50~(2) (2002)
  281--289.

\bibitem{Andrieu03}
C.~Andrieu, A.~Doucet, Online expectation-maximization type algorithms for
  parameter estimation in general state space models, in: 2003 IEEE
  International Conference on Acoustics, Speech, and Signal Processing
  (ICASSP), Vol.~6, IEEE, 2003, pp. VI--69.

\bibitem{Ding10}
J.~Ding, Y.~Shi, H.~Wang, F.~Ding, A modified stochastic gradient based
  parameter estimation algorithm for dual-rate sampled-data systems, Digital
  Signal Processing 20~(4) (2010) 1238--1247.

\bibitem{Ding13}
F.~Ding, Y.~Gu, Performance analysis of the auxiliary model-based stochastic
  gradient parameter estimation algorithm for state-space systems with one-step
  state delay, Circuits, Systems, and Signal Processing 32~(2) (2013) 585--599.

\bibitem{Kokkala15}
J.~Kokkala, S.~S{\"a}rkk{\"a}, Combining particle {MCMC} with
  {R}ao-{B}lackwellized {M}onte {C}arlo data association for parameter
  estimation in multiple target tracking, Digital Signal Processing 47 (2015)
  84--95.

\bibitem{Andrieu10}
C.~Andrieu, A.~Doucet, R.~Holenstein, Particle {M}arkov chain {M}onte {C}arlo
  methods, Journal of the Royal Statistical Society B 72 (2010) 269--342.

\bibitem{Koblents15}
E.~Koblents, J.~M\'{\i}guez, A population monte carlo scheme with transformed
  weights and its application to stochastic kinetic models, Statistics and
  Computing 25~(2) (2015) 407--425.

\bibitem{Crisan15}
D.~Crisan, J.~Miguez, Nested particle filters for online parameter estimation
  in discrete-time state-space markov models, arXiv 1308.1883v3 [stat.CO].

\bibitem{Kantas15}
N.~Kantas, A.~Doucet, S.~S. Singh, J.~M. Maciejowski, N.~Chopin, On particle
  methods for parameter estimation in state-space models, Statistical Science
  30 (2015) 328--351.

\bibitem{Olsson11}
J.~Olsson, T.~Ryden, {R}ao-{B}lackwellization of particle {M}arkov chain
  {M}onte {C}arlo methods using forward filtering backward sampling, IEEE
  Transactions on Signal Processing 59~(10) (2011) 4606--4619.

\bibitem{Vu14}
T.~Vu, B.-N. Vo, R.~Evans, A particle marginal {M}etropolis-{H}astings
  multi-target tracker, IEEE Transactions on Signal Processing 62~(15) (2014)
  3953--3964.

\bibitem{Kwon16}
J.~Kwon, R.~Dragon, L.~Van~Gool, Joint tracking and ground plane estimation,
  IEEE Signal Processing Letters 23~(11) (2016) 1514--1517.

\bibitem{Ala16}
J.~Ala-Luhtala, N.~Whiteley, K.~Heine, R.~Pich{\'e}, An introduction to twisted
  particle filters and parameter estimation in non-linear state-space models,
  IEEE Transactions on Signal Processing 64~(18) (2016) 4875--4890.

\bibitem{Fitzgerald01}
W.~J. Fitzgerald, {M}arkov chain {M}onte {C}arlo methods with applications to
  signal processing, Signal Processing 81~(1) (2001) 3--18.

\bibitem{Gordon93}
N.~Gordon, D.~Salmond, A.~F.~M. Smith, Novel approach to nonlinear and
  non-{G}aussian {B}ayesian state estimation, IEE Proceedings-F 140~(2) (1993)
  107--113.

\bibitem{Doucet01b}
A.~Doucet, N.~de~Freitas, N.~Gordon (Eds.), Sequential {M}onte {C}arlo Methods
  in Practice, Springer, New York (USA), 2001.

\bibitem{Doucet00}
A.~Doucet, S.~Godsill, C.~Andrieu, On sequential {M}onte {C}arlo {S}ampling
  methods for {B}ayesian filtering, Statistics and Computing 10~(3) (2000)
  197--208.

\bibitem{Djuric03}
P.~M. Djuri\'c, J.~H. Kotecha, J.~Zhang, Y.~Huang, T.~Ghirmai, M.~F. Bugallo,
  J.~M\'{\i}guez, Particle filtering, IEEE Signal Processing Magazine 20~(5)
  (2003) 19--38.

\bibitem{Cappe07}
O.~Capp{\'e}, S.~J. Godsill, E.~Moulines, An overview of existing methods and
  recent advances in sequential {M}onte {C}arlo, Proceedings of the IEEE 95~(5)
  (2007) 899--924.

\bibitem{Robert04}
C.~P. Robert, G.~Casella, {M}onte {C}arlo Statistical Methods, Springer, 2004.

\bibitem{Cappe04}
O.~Capp\'e, A.~Gullin, J.~M. Marin, C.~P. Robert, Population monte carlo,
  Journal of Computational and Graphical Statistics 13~(4) (2004) 907--929.

\bibitem{Chopin12}
N.~Chopin, P.~E. Jacob, O.~Papaspiliopoulos, {SMC2}: an efficient algorithm for
  sequential analysis of state space models, Journal of the Royal Statistical
  Society: Series B (Statistical Methodology).

\bibitem{Hong10}
M.~Hong, M.~F. Bugallo, P.~M. Djuric, Joint model selection and parameter
  estimation by population {M}onte {C}arlo simulation, IEEE Journal of Selected
  Topics in Signal Processing 4~(3) (2010) 526--539.

\bibitem{Martino15}
L.~Martino, V.~Elvira, D.~Luengo, J.~Corander, An adaptive population
  importance sampler: Learning from uncertainty, IEEE Transactions on Signal
  Processing 63~(16) (2015) 4422--4437.

\bibitem{Bugallo15}
M.~F. Bugallo, L.~Martino, J.~Corander, Adaptive importance sampling in signal
  processing, Digital Signal Processing 47 (2015) 36--49.

\bibitem{Elvira17}
V.~Elvira, L.~Martino, D.~Luengo, M.~F. Bugallo, Improving population monte
  carlo: Alternative weighting and resampling schemes, Signal Processing 131
  (2017) 77--91.

\bibitem{Chopin02}
N.~Chopin, A sequential particle filter method for static models, Biometrika
  89~(3) (2002) 539--552.

\bibitem{DelMoral06}
P.~Del~Moral, A.~Doucet, A.~Jasra, Sequential {M}onte {C}arlo samplers, Journal
  of the Royal Statistical Society: Series B (Statistical Methodology) 68~(3)
  (2006) 411--436.

\bibitem{Kong94}
A.~Kong, J.~S. Liu, W.~H. Wong, Sequential imputations and {B}ayesian missing
  data problems, Journal of the American Statistical Association 9 (1994)
  278--288.

\bibitem{Elvira15b}
V.~Elvira, L.~Martino, D.~Luengo, M.~F. Bugallo, Efficient multiple importance
  sampling estimators, IEEE Signal Processing Letters 22~(10) (2015)
  1757--1761.

\bibitem{Bain08}
A.~Bain, D.~Crisan, Fundamentals of Stochastic Filtering, Springer, 2008.

\bibitem{Anderson79}
B.~D.~O. Anderson, J.~B. Moore, Optimal Filtering, Englewood Cliffs, 1979.

\bibitem{Kitagawa96}
G.~Kitagawa, Monte {C}arlo filter and smoother for non-{G}aussian nonlinear
  state-space models, J. Comput. Graph. Statist. 1 (1996) 1--25.

\bibitem{Douc05}
R.~Douc, O.~Capp\'e, E.~Moulines, Comparison of resampling schemes for particle
  filtering, in: Proceedings of the 4${}^{th}$ International Symposium on Image
  and Signal Processing and Analysis, 2005, pp. 64--69.

\bibitem{DelMoral04}
P.~{Del Moral}, {F}eynman-{K}ac Formulae: Genealogical and Interacting Particle
  Systems with Applications, Springer, 2004.

\bibitem{Miguez13b}
J.~M{\'{\i}}guez, D.~Crisan, P.~M. Djuri\'c, On the convergence of two
  sequential {M}onte {C}arlo methods for maximum a posteriori sequence
  estimation and stochastic global optimization, Statistics and Computing
  23~(1) (2013) 91--107.

\bibitem{Andrieu09}
C.~Andrieu, G.~Roberts, The pseudo-marginal approach for efficient {M}onte
  {C}arlo computations, Annals of Statistics 37 (2009) 697--725.

\bibitem{Doucet01}
A.~Doucet, N.~de~Freitas, N.~Gordon, An introduction to sequential {M}onte
  {C}arlo methods, in: A.~Doucet, N.~de~Freitas, N.~Gordon (Eds.), Sequential
  {M}onte {C}arlo Methods in Practice, Springer, 2001, Ch.~1, pp. 4--14.

\bibitem{Crisan15par}
D.~Crisan, J.~Miguez, G.~R\'ios, A simple scheme for the parallelisation of
  particle filters and its application to the tracking of complex stochastic
  systems, arXiv arXiv:1407.8071v2 [stat.CO].

\bibitem{Koblents16}
E.~Koblents, J.~Miguez, M.~A. Rodriguez, A.~M. Schmidt, A nonlinear population
  {M}onte {C}arlo scheme for the {B}ayesian estimation of parameters of
  $\alpha$-stable distributions, Computational Statistics and Data Analysis 95
  (2016) 57--74.

\bibitem{Crisan14a}
D.~Crisan, J.~Miguez, Particle-kernel estimation of the filter density in
  state-space models, Bernoulli 20~(4) (2014) 1879--1929.

\bibitem{Farin13}
G.~Farin, D.~Hansford, Practical linear algebra: A geometry toolbox, CRC Press,
  2013.

\end{thebibliography}
